\documentclass[twocolumn,tighten]{aastex63}

\usepackage{color}
\usepackage{comment}
\usepackage{amsmath}
\usepackage{amssymb}
\usepackage{comment}
\usepackage{xcolor}

\usepackage{CJK} 

\shorttitle{Magellanic Stellar Debris}
\shortauthors{}

\begin{document}
\begin{CJK*}{UTF8}{gbsn}

\title{Discovery of Magellanic Stellar Debris in the H3 Survey}


\author[0000-0002-5177-727X]{Dennis Zaritsky}
\affiliation{Steward Observatory, University of Arizona, 933 North Cherry Avenue, Tucson, AZ 85721-0065, USA}

\author[0000-0002-1590-8551]{Charlie Conroy}
\affiliation{Center for Astrophysics $|$ Harvard \& Smithsonian, 60 Garden Street, Cambridge, MA 02138, USA}

\author[0000-0003-3997-5705]{Rohan P. Naidu}
\affiliation{Center for Astrophysics $|$ Harvard \& Smithsonian, 60 Garden Street, Cambridge, MA 02138, USA}

\author[0000-0002-1617-8917]{Phillip A. Cargile}
\affiliation{Center for Astrophysics $|$ Harvard \& Smithsonian, 60 Garden Street, Cambridge, MA 02138, USA}

\author[0000-0002-1129-1873]{Mary Putman}
\affiliation{Department of Astronomy, Columbia University, 550 West 120th Street, New York, NY 10027, USA}

\author[0000-0003-0715-2173]{Gurtina Besla}
\affiliation{Steward Observatory, University of Arizona, 933 North Cherry Avenue, Tucson, AZ 85721-0065, USA}

\author[0000-0002-7846-9787]{Ana Bonaca}
\affiliation{Center for Astrophysics $|$ Harvard \& Smithsonian, 60 Garden Street, Cambridge, MA 02138, USA}

\author[0000-0003-2352-3202]{Nelson Caldwell}
\affiliation{Center for Astrophysics $|$ Harvard \& Smithsonian, 60 Garden Street, Cambridge, MA 02138, USA}

\author[0000-0002-6800-5778]{Jiwon Jesse Han}
\affiliation{Center for Astrophysics $|$ Harvard \& Smithsonian, 60 Garden Street, Cambridge, MA 02138, USA}

\author[0000-0002-9280-7594]{Benjamin D. Johnson}
\affiliation{Center for Astrophysics $|$ Harvard \& Smithsonian, 60 Garden Street, Cambridge, MA 02138, USA}

\author[0000-0003-2573-9832]{Joshua S. Speagle}
\altaffiliation{Banting Fellow}
\affiliation{Center for Astrophysics $|$ Harvard \& Smithsonian, 60 Garden Street, Cambridge, MA 02138, USA}
\affiliation{Department of Statistical Sciences, University of Toronto, Toronto, M5S 3G3, Canada}
\affiliation{Dunlap Institute for Astronomy \& Astrophysics, University of Toronto, Toronto, M5S 3H4, Canada}
\affiliation{David A. Dunlap Department of Astronomy \& Astrophysics, University of Toronto, Toronto, M5S 3H4, Canada}

\author[0000-0001-5082-9536]{Yuan-Sen Ting (丁源森)}
\altaffiliation{Hubble Fellow}
\affiliation{Institute for Advanced Study, Princeton, NJ 08540, USA}
\affiliation{Department of Astrophysical Sciences, Princeton University, Princeton, NJ 08544, USA}
\affiliation{Observatories of the Carnegie Institution of Washington, 813 Santa Barbara Street, Pasadena, CA 91101, USA}
\affiliation{Research School of Astronomy and Astrophysics, Mount Stromlo Observatory, Cotter Road, Weston Creek, ACT 2611, Canberra, Australia}

\email{dennis.zaritsky@gmail.com}


\begin{abstract}
We report the discovery of 15 stars in the H3 survey that lie, in projection, near the tip of the trailing gaseous Magellanic Stream (MS). The stars have Galactocentric velocities $< -155$ km s$^{-1}$, Galactocentric distances of $\approx 40$ to 80 kpc (increasing along the MS), and [Fe/H] consistent with that of stars in the Small Magellanic Cloud. These 15 stars comprise $94$\% (15 of 16) of the H3 observed stars to date that have $R_{GAL} > 37.5$ kpc, $-$350 km s$^{-1} < V_{GSR} < -155$ km s$^{-1}$, and are not associated with the Sagittarius Stream. They represent a unique portion of the Milky Way's outer halo phase space distribution function and confirm that unrelaxed structure is detectable even at radii where H3 includes only a few hundred stars. Due to their statistical excess, their close association with the MS and H{\small I} compact clouds in the same region, both in position and velocity space, and their plausible correspondence with tidal debris in a published simulation, we identify these stars as debris of past Magellanic Cloud encounters. These stars are evidence for a stellar component of the tidal debris field far from the Clouds themselves and provide unique constraints on the interaction. 

\end{abstract}





\section{Introduction}

The gas trailing the Large and Small Magellanic Clouds across $\sim$ 150$^\circ$ on the sky, referred to as the Magellanic Stream \citep{mathewson}, testifies to a complex dynamical interaction both between the Clouds themselves and the Milky Way. In response to the discovery of this gas, two families of origin stories arose: those where tidal forces are predominantly responsible for removing gas from either or both Clouds \citep[e.g.,][]{fujimoto,lin,murai} and those where hydrodynamical forces are \citep[e.g.,][]{meurer,moore}. 

Both families have faced significant challenges \citep[for a recent review see][]{elena}. With our focus here on stars, we consider the tidal models more closely. These have faced two principal difficulties. First, the initial observations of the Magellanic Stream (MS) only identified a trailing arm. Tidal interactions are predicted to create both leading and trailing arms. This discrepancy was addressed with the discovery of the leading `arm' \citep{putman} and with simulations that identified scenarios where the leading arm is not as prominent as the trailing arm \citep{besla12,lucchini} even as other studies questioned this interpretation \citep{tepper-garcia}. 
Second, tidal interactions are generally expected to remove both gas and stars from the Clouds and yet searches for stars associated with the MS usually produced null results \citep{recillas,brueck,raja}. 

There are a few tantalizing exceptions to the empirical absence of tidal stars. One such exception is the work of \cite{belokurov}, who identified clumps of blue horizontal branch stars out to at least 30$^\circ$ from the Clouds, some of which share similar proper motions with the Clouds and some of which are coincident on the sky with gas in the MS. The nature of these apparent clumps remains an open question.
A second is the work of \cite{mackey}, who identify a stellar feature near the disk of the LMC that could possibly be the `headwaters' of a stellar stream and is also seen by \cite{deason1}. A third, is an excess of BHB stars seen in one of the
\cite{deason18} survey fields, which they hypothesize may be Magellanic tidal debris, although they cannot rule out field-to-field variations as the cause of the excess.
Finally, increasingly sensitive surveys continue to find stars belonging to the Clouds at larger and larger radii, including structures that could be the result of tidal interactions \citep{nidever19a}.

The confirmation of stellar  tidal debris far beyond the tidal radii of the Clouds would have immediate ramifications on the question of the origin of the MS. First, it would  validate the tidal origin scenario. Second, it would enable distance measurements to some portion of the tidal debris field. Gas associated with those stars could be spatially offset if hydrodynamical forces also play a role \citep[see][for a recent and extensive treatment]{tepper-garcia}, but, if that offset is modest, the distance measurement would enable us to derive gas masses along the MS, which is key to understanding the origin and evolution of this component. Third, distance constraints to any component of the tidal debris field will help distinguish among different orbital models of the Magellanic system. Finally, if one can complement the distance measurements of the stars with distance measurements to associated gas, then one can probe the properties of the Milky Way's hot coronal gas and the physical interplay between that gas and the local radiation field \citep[cf.,][]{bregman,weiner,bland-hawthorn,murali,gnat10}. Because tidal and hydrodynamical effects are both likely to be important in sculpting the tidal debris, identifying and measuring the properties of both the gaseous and stellar debris will be key in untangling the effects of the two processes. 

\begin{deluxetable*}{llrrrrr}
\tablewidth{0pt}
\label{tab:stars}
\tablecaption{Selected H3 Stars\tablenote{Star 111881656 is reclassified as a Sgr member using the \cite{johnson} criteria when $R_{GAL,NP}$ is used in place of $R_{GAL}$. This star is the outlier relative to the MS in Figures \ref{fig:money} and \ref{fig:sampling}}}
\tablehead{  \colhead{GAIA ID} & \colhead{H3 ID} &  \colhead{RA} & \colhead{Dec} & \colhead{R$_{GAL}$} & \colhead{R$_{GAL,NP }$}&\colhead{$V_{GSR}$} \\
&&&& \colhead{[kpc]} & \colhead{[kpc]}&\colhead{[km s$^{-1}$]} 
}
\startdata
2696592708532990464&111881656&328.0350512&  4.254902&  56$\pm$5 & 67$\pm$9   &$-$211.6$\pm$0.5\\
2714515130318602112&117592880&347.2415753& 10.341434&  77$\pm$3 &79$\pm$3  &$-$157.7$\pm$0.3\\
2812229896910034176&117763491&348.1424296& 12.910014&  63$\pm$3 &77$\pm$7  &$-$163.0$\pm$0.4\\
2630501957940493184&118360373&348.3172866& $-$8.054611&  51$\pm$2 &57$\pm$6  &$-$189.0$\pm$0.3\\
2633573237514082304&119163672&349.5990965& $-$5.356268&  40$\pm$4 &49$\pm$7  &$-$220.7$\pm$0.6\\
2635522259313314560&119332861&346.3077439& $-$4.986723&  53$\pm$7 &75$\pm$9  &$-$230.2$\pm$0.9\\
2610348528278657664&119511723&342.7162418& $-$7.813899&  41$\pm$1 &55$\pm$3  &$-$179.2$\pm$0.2\\
2433738824527479168&119718763&354.4038651&$-$11.362019&  38$\pm$2 &45$\pm$7  &$-$194.7$\pm$0.4\\
2433795548159999488&119719901&354.8523545&$-$11.153222&  41$\pm$3 &55$\pm$6  &$-$201.2$\pm$0.4\\
2422797172003289088&119829926&359.9671116&$-$10.730789&  47$\pm$2 &60$\pm$7  &$-$203.2$\pm$0.3\\
2422487281523040384&119830529&359.9326760&$-$11.184418&  48$\pm$4 &59$\pm$7  &$-$184.9$\pm$0.4\\
2406230521069229312&120102214&348.3230413&$-$17.049562&  51$\pm$2 &  55$\pm$3  &$-$215.8$\pm$0.2\\
2408253858687755264&120125054&352.4714532&$-$13.831915&  54$\pm$11&72$\pm$9  &$-$179.3$\pm$1.0\\
2436871844255646336&120442318&351.6534275&$-$10.711871&  39$\pm$3 &57$\pm$7  &$-$206.2$\pm$0.6\\
2419420915391260288&120967601&358.7294761&$-$14.353803&  45$\pm$3 &57$\pm$7  &$-$180.6$\pm$0.5\\
\enddata
\end{deluxetable*}

We present the serendipitous discovery of halo stars that are projected near the tip of the trailing MS and have similar kinematics to gas clouds found in the same area. Because the `smooth' halo of the Milky Way at these distances ($>$ 40 kpc) has kinematics that are distinct from this small sample of 15 stars, the contrast between these stars and halo stars is remarkable. We describe the general characteristics of the H3 survey in \S \ref{sec:data}, and our selection of this particular set of stars and what characteristics lead us to classify them as Magellanic Clouds tidal debris in \S \ref{sec:discussion}. In that same section, we close with a brief discussion of possible inferences that might made from this population of stars and include a comparison to a previous numerical simulation of the stellar tidal debris field.

\section{Data}
\label{sec:data}

The H3 survey provides high-resolution spectroscopy of likely halo stars in a sparse grid covering roughly 15,000 square degrees \citep{conroy}. Likely halo stars are identified by requiring that stars in high Galactic latitude fields  ($\vert b \vert>30^{\circ}$ and Dec. $>-20^{\circ}$) satisfy
$15<r<18$ and by placing constraints on the parallax that defines a lower distance bound. In practice, the latter requirement translated to a requirement on the Gaia parallax, $\varpi$, that changed slightly during the survey. At first we required $\varpi - 2\sigma_\varpi<0.5$, which we later changed to $\varpi<0.4$. The change has no effect in this study because we are focusing on the most distant stars in H3.

We obtain the spectra using the fiber-fed Hectochelle spectrograph on the MMT in a configuration that produces spectra with resolution values of  32,000 from $5150$ to $5300$\AA. 
From these spectra, H3 will produce a catalog of stellar
parameters and spectrophotometric distances for $\approx$ 200,000 stars when completed.
The procedure we use in determining stellar parameters and distance estimates was developed and presented by \cite{cargile2020}. These quantities are derived using a `Galactic' prior, whose influence on the parameters of interest for our stars is to bias the derived Galactocentric distances, $R_{GAL}$, downward. Re-fitting with a flat prior, results in a derived distance, $R_{GAL,NP}$, that increase by $\lesssim$ 20\%. Nevertheless, for consistency with the available catalog, we focus our discussion using the original distances here. This distance uncertainty does not affect any of our conclusions, although the revised distances are in less tension with models that we describe in \S3. 
In contrast, the values of $V_{GSR}$ are quite precise given that $V_{RAD}$ is measured to $\sim$ 1 km sec$^{-1}$ precision and the conversion to $V_{GSR}$ depends only on the Sun's position in the Galaxy. As the survey has progressed toward the full sample, a number of studies using the data available at the time have been published that address different scientific questions  \citep{conroy_metals,bonaca,zar20,naidu,johnson}. 

From the available set of observed and analyzed H3 stars ( rcat\_V3.0\_MSG.fits), about 136,000, we select those that had no spectral fitting problems (${\rm FLAG} = 0$), spectral signal-to-noise ratio (SNR) per pixel $>$ 3, and were not previously identified as associated with the Sagittarius stream \citep[${\rm FLAG}_{\rm SGR} = 0$;][]{johnson}.
After these cuts, we were left with about 95,000 stars.

\begin{figure*}
\includegraphics[width=1.0\textwidth]{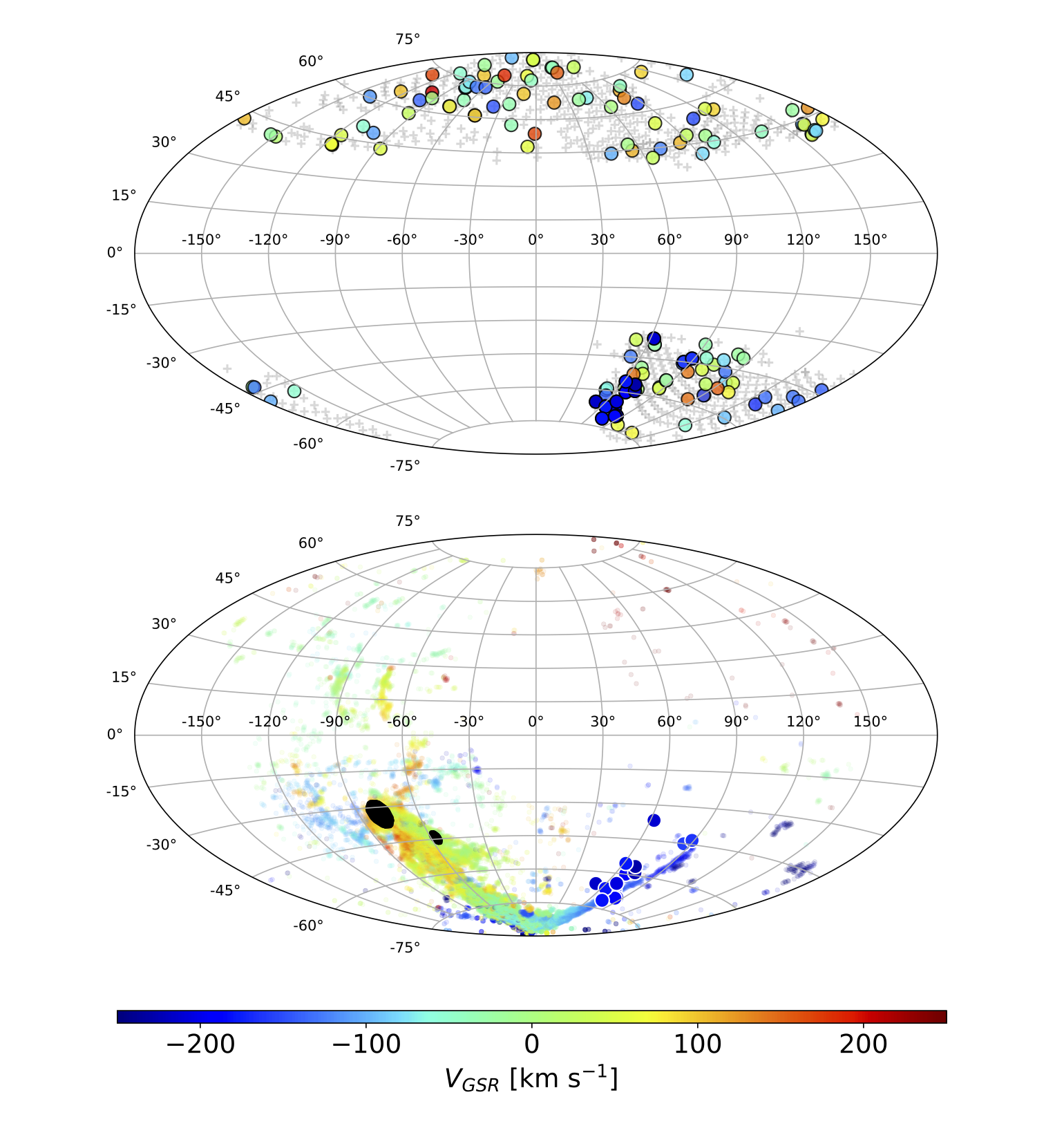}
\caption{(Upper Panel) H3 halo stars to date at Galactocentric distances $>$ 37.5 kpc, excluding stars identified as Sgr members or failed quality cuts (see text), are shown as colored circles, coded by V$_{GSR}$. The fields observed so far in H3 are shown in grey plus signs. (Lower Panel) Our subsample of H3 stars, those from the upper panel with $-350 <$ V$_{GSR} < -155$ km s$^{-1}$, are compared to the gaseous Magellanic Stream, which is color coded to the same scale in V$_{GSR}$. The LMC and SMC are shown as black ellipses. }
\label{fig:money}
\end{figure*}

\section{Results}
\label{sec:discussion}

\subsection{Identification of  Substructure in the Outer Halo}

During a search for stars that could be used to measure absorption by the MS, and thereby constrain the distance to the MS, we found a population of stars that roughly shares the velocity of the MS. Intrigued by the possibility that these might be associated with the MS, we selected  stars in H3 at large Galactocentric distance, $>$ 40 kpc,  with comparable Galactocentric velocities to the MS in this region of sky, $-$350  km s$^{-1} <$ V$_{GSR} < -155$ km s$^{-1}$. We found a tight cluster of 12 stars in the sky. Slightly adjusting the distance to maximize the number of stars in this clump and minimize contamination (e.g., stars at positive Galactic latitudes), we settled on the criteria $R_{GAL} > 37.5$ kpc to select 16 stars. We remove from consideration one star from this group that is projected in an area of sky more closely related to Sgr debris (at similar latitude but at $l > 150^\circ$).  
The distribution of the remaining 15 
selected stars within the H3 footprint is manifestly not random (Figure \ref{fig:money}) and we provide the particulars of those stars in Table \ref{tab:stars}.

Before proceeding to make the case that the 15 stars are likely to be associated with the Magellanic Cloud tidal debris, we pause to emphasize an important result and its implications. Even within the fairly small population of known outer-halo stars in H3, there is a strong, previously-unknown signature of unrelaxed substructure. Such substructure poses a difficult challenge for any analysis of the halo that presumes a relatively smooth distribution function. For example, analyses of the escape velocity, which aim to constrain the mass of the Milky Way, often adopt simple analytic expressions \citep[cf.,][]{piffl,williams,deason} for the tail of the velocity distribution, although they do acknowledge and attempt to account for substructure \citep{piffl,grand2019}. The current finding empirically illustrates this problem exists out at as far out as H3 can probe with at least a few hundred stars. Our previous analysis of the halo at smaller distances shows the prevalence of substructure throughout the halo \citep{naidu}.

\subsection{Connection to Magellanic Stream and Debris Field}

In the bottom panel of Figure \ref{fig:money}, we compare the distribution of our selected 15 H3 stars to that of the gas in the MS as presented by \cite{nidever}.
The MS passes close to the stars in our sample. As seen in the figure, the Galactocentric velocities, color coded in the range from $-250 < V_{GSR}/({\rm km\  s}^{-1}) < 250$, are also similar between gas and stars. Although the stars and MS are nearly coincident on the sky and in velocity, we do not have distance measurements for the MS. As such, we cannot determine whether the stars and gas are truly co-spatial. A further complication in associating the H3 stars with the MS, or any other gas in the area, is that because of the expected hydrodynamical drag experienced by gas in the Milky Way halo the positions on the sky, velocities, or distances of the two populations could differ even if they share an origin \citep[see][for a discussion of such an offset in the leading arm]{nidever19b}.

\begin{figure}
\includegraphics[width=0.5\textwidth]{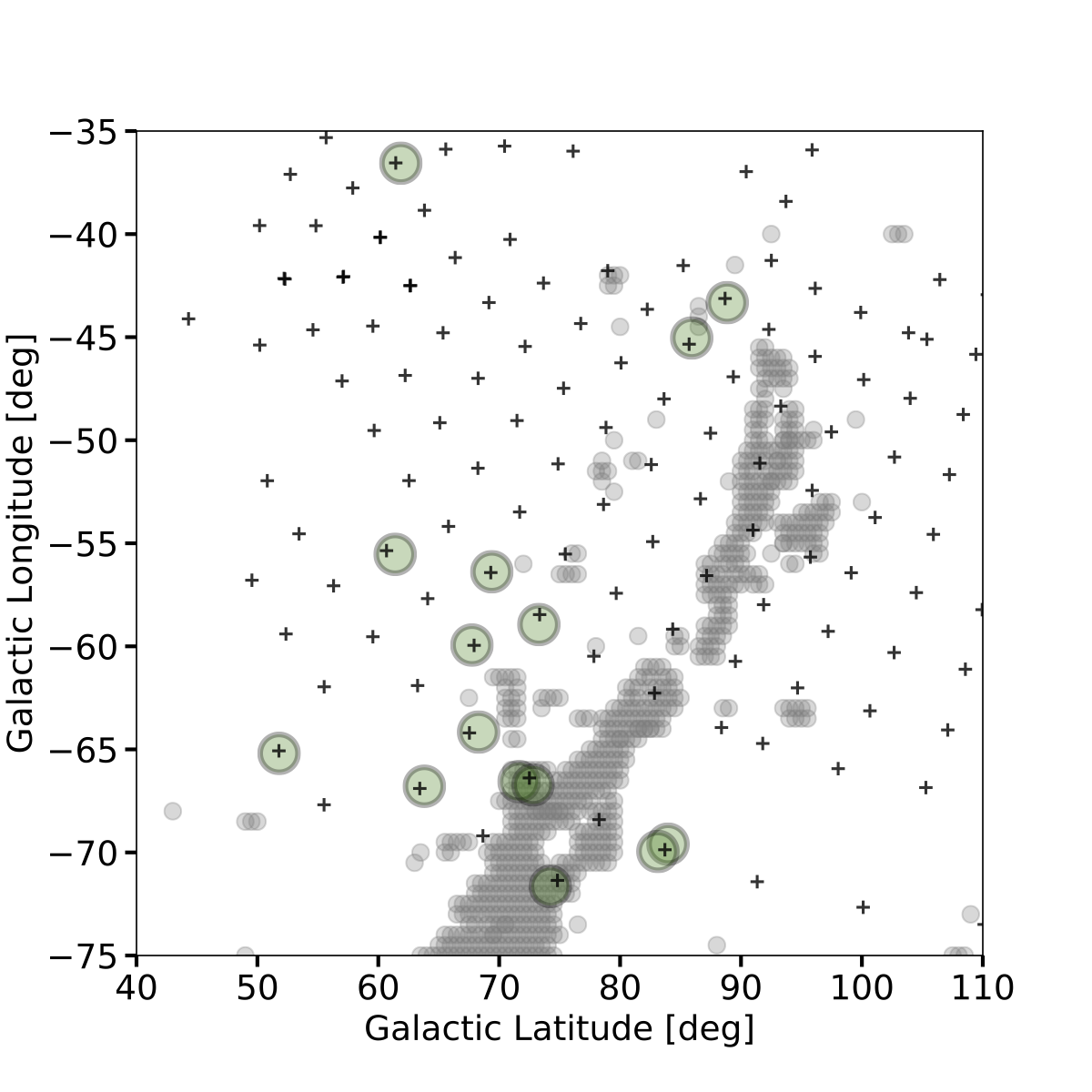}
\caption{Our H3 selected stars (large green circles) and H3 field centers observed to date (plus signs) in the area near the tip of the Magellanic Stream (MS). The light grey small circles map the gaseous MS in this region from \cite{nidever}.}
\label{fig:sampling}
\end{figure}

\begin{figure*}
\includegraphics[width=1.0\textwidth]{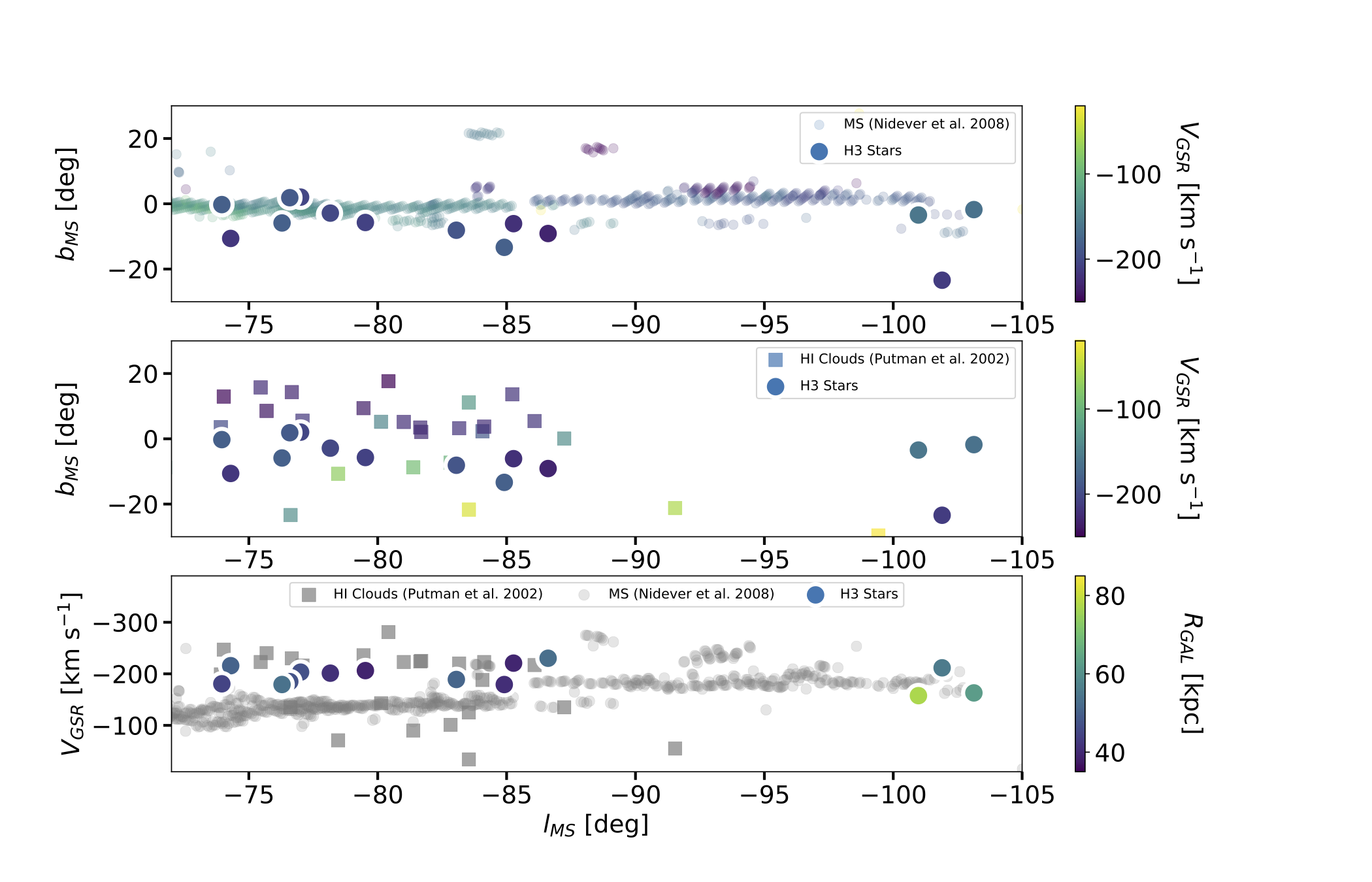}
\caption{Distribution of stars and gas in the Magellanic Stream tip region. In the upper panel we plot the distribution of stars (large circles), the MS (from \cite{nidever} in small circles), while in the middle panel we plot the same stars and compact clouds 
from \cite{putman02} as squares. We present the data 
in Magellanic Stream coordinates from \cite{nidever}.
All symbols are color coded according to V$_{GSR}$ in the top two panels. 
In the lowest panel we plot the same objects as in the top two panels in different coordinates (V$_{GSR}$ vs. Magellanic Stream longitude). Symbols remain the same, but the color coding reflects $R_{GAL}$ only for the stars (otherwise symbols are gray because distances to the gas are unknown).  }
\label{fig:besla}
\end{figure*}

Considering this caveat, we examine the correspondence between gas and stars more closely in Figures \ref{fig:sampling} and \ref{fig:besla}. Beginning with Figure \ref{fig:sampling}, we see that the MS appears to consist of two filaments, one of which is significantly more prominent than the other. The existence of apparently intertwined filaments, although again we do not know their distances, at various locations along the MS is long-established \citep{cohen,morras,putman03,nidever} and this morphology persists to the tail of the MS as shown. Our H3 stars are visually more closely related to the weaker filament, but more stars are needed to confirm this suggestion. We caution that H3 samples the sky sparsely, so an absence of H3 sources in any particular region may not reflect a true scarcity. Nevertheless, H3 has sampled within the stronger of the two filaments and found no stars that are unambiguously projected solely on that filament (Figure \ref{fig:sampling}). 

In Figure \ref{fig:besla} we extend our examination of the association in position and velocity space between the selected H3 stars, the MS, and compact gas clouds identified by \cite{putman02} 
in this same region. In that work,
if the H{\small I} emission is continuous in position and velocity, as it is in large sections of the main body of the Magellanic Stream, it is cataloged as a single cloud.  Therefore, the individual clouds featured in Figure \ref{fig:besla} are predominantly smaller, discrete clouds.

The velocities of the H3 stars are offset by several tens of km s$^{-1}$ from those of the MS, particularly at southerly latitudes. This result may be, at least in part, due to our velocity selection. The MS in this region of the sky has a mean velocity $>$ $-$155 km s$^{-1}$, a range which we excluded in our selection. To examine the impact of this exclusion, we raise the cutoff velocity to $-100$ km s$^{-1}$ from $-155$ km s$^{-1}$. We find four additional stars that fall within the region plotted in Figures \ref{fig:sampling} and \ref{fig:besla}, but also find a larger number of selected stars  distributed throughout the H3 footprint, making it difficult to confidently assert that the new stars found near the MS are not interlopers. The lack of an overdensity near the MS for stars with $-155 < V_{GSR}/$(km s$^{-1}) < -100$ indicates that the stars we discuss here are not the highly 
negative velocity tail of a large-scale asymmetry in the halo, such as that induced by the proposed Magellan Cloud wake \citep{gavarito}. We conclude that:
\begin{enumerate}
    \item We are unable to fully sample the possible velocity range of stars in this population, so detailed comparisons of the mean velocity or velocity dispersion of the stellar component to the MS or compact gas clouds are compromised. Nevertheless, this is a kinematically cold component with a dispersion in $V_{GSR}$ of 20 km sec$^{-1}$.
    \item We are fortunate that at least some of the Magellanic tidal debris falls within the H3 footprint and that the velocity distribution of the debris in this area of sky is the most dissimilar from the general Milky Way halo.
\end{enumerate}
The latter enabled us to distinguish even a small population of debris stars. Extending a spectroscopic survey along the MS toward the Magellanic Clouds will require a much larger number of survey stars to confirm an excess along the MS because those stars will be less distinct in velocity from the general halo population.

In Figure \ref{fig:besla} we see that the stars are offset in velocity from the MS at comparable MS longitude, $l_{MS}$, and offset in position relative to the compact clouds with which they are best matched in velocity. The situation becomes even more complicated when we examine the lowest panel in the figure, where the MS bifurcates into two pieces with velocities that differ by $\sim$ 50 km s$^{-1}$ and the compact clouds in the left half of the plotted longitude range are at velocities that are both smaller and larger than those of the MS. The stars are a closer match in velocity to the \cite{putman02} clouds (although recall that we exclude stars with V$_{GSR}>-155$ km s$^{-1}$) and to the MS at more negative MS longitude, but are projected closer on the sky to the MS 
with which the velocity offset is very large $\sim 150$ km s$^{-1}$. The stars also appear likely to be physically more closely associated with the compact clouds in that their distributions in $l_{MS}$ both appear to decline sharply for $l_{MS} \lesssim -87^\circ$. Associating the stars with either the MS or a population of compact clouds requires accepting either a position or velocity offset, and is therefore uncertain.

In the lowest panel of Figure \ref{fig:besla} we color code the stars by Galactocentric distance. There is a distance gradient in the direction of decreasing MS longitude with stellar distances ranging from 40 to 60 kpc at one end and 65 to 80 kpc for stars projected near the MS tip (recall that re-fitting the spectra with flat priors increases these distances by $\lesssim$ 20\%). H3 overall has few stars at the upper limit of this range and beyond, so we cannot determine whether we are simply seeing the near tail of a distribution that extends far along the line of sight or whether this is a physically thin structure of stars. We do not find an excess of stars at these velocities in this region of the sky when we consider $R_{GAL} < 37.5$ kpc. We conclude that this population does not extend closer to us than these distances suggest.
The distance to Magellanic Cloud tidal debris, particularly of stars, which are not affected by drag in the Galaxy's hot corona, is potentially a highly discriminating constraint on complex interaction models \citep[e.g.,][]{gomez}.

\subsection{Origin Story}

The origin of the MS, the compact clouds, and now of these stars too, remain open questions. \cite{nidever} argued against the previous prevailing hypothesis that the MS comes mostly from the Small Magellanic Cloud (SMC). They concluded that one of the two gaseous filaments originates from the Large Magellanic Cloud (LMC). This understanding has been further supported by metallicity measurements \citep{richter}, but the higher metallicity is seen only in one sightline that is relatively close to the LMC and ram pressure stripping of gas from the LMC \citep{salem} could contaminate tidal material. A non-LMC origin is supported by the lower abundance measurements of the gas in one filament \citep{fox}, although there are only five sightlines and in several cases it is unclear which of the two filaments is being probed. Furthermore, as noted by \cite{fox}, the interpretation of the abundances is complicated by the chemical enrichment history of the responsible  galaxy since the time when the gas was lost, by a possible abundance gradient in the responsible galaxy, and by subsequent mixing of the gas with presumably more pristine halo gas. In support of an SMC origin for at least some of the tidal material, \cite{for} concluded on the basis of kinematic arguments that the origin of the compact clouds they observed is the SMC. 

Although ascribing a common origin to these stars and the MS, or compact clouds, is appealing, there are two principal characteristics of the stars that complicate such an interpretation.
First, the distances to these stars are a roughly factor of two to three smaller than the distance to the MS at this location on the sky suggested by extensive modeling \citep{besla12,gomez,pardy}.
Second, the angular momenta of the stars and the Magellanic Clouds are quite different, most noticeably the x-component of either Magellanic Cloud is $< -10,000$ km s$^{-1}$ kpc, while that of the stars tend to be $>$ 0 km s$^{-1}$ kpc. 

To examine these issues further, we avail ourselves of the published simulation of \cite{besla13}, who present positions and velocities for SMC stellar tidal debris. Our aim is to determine if it is plausible that the H3 stars are part of such a debris field. There exist other simulations, which differ both in initial conditions and outcomes \citep[e.g.,][]{pardy,tepper-garcia}, but our goal here is only to determine if these stars have a plausible origin in the interaction. A full comparison between these observations and simulations, which may provide insight into the validity of the model assumptions and initial conditions, would require exploring the simulation parameter space and is therefore beyond the scope of this work.

\begin{figure}
\includegraphics[width=0.5\textwidth]{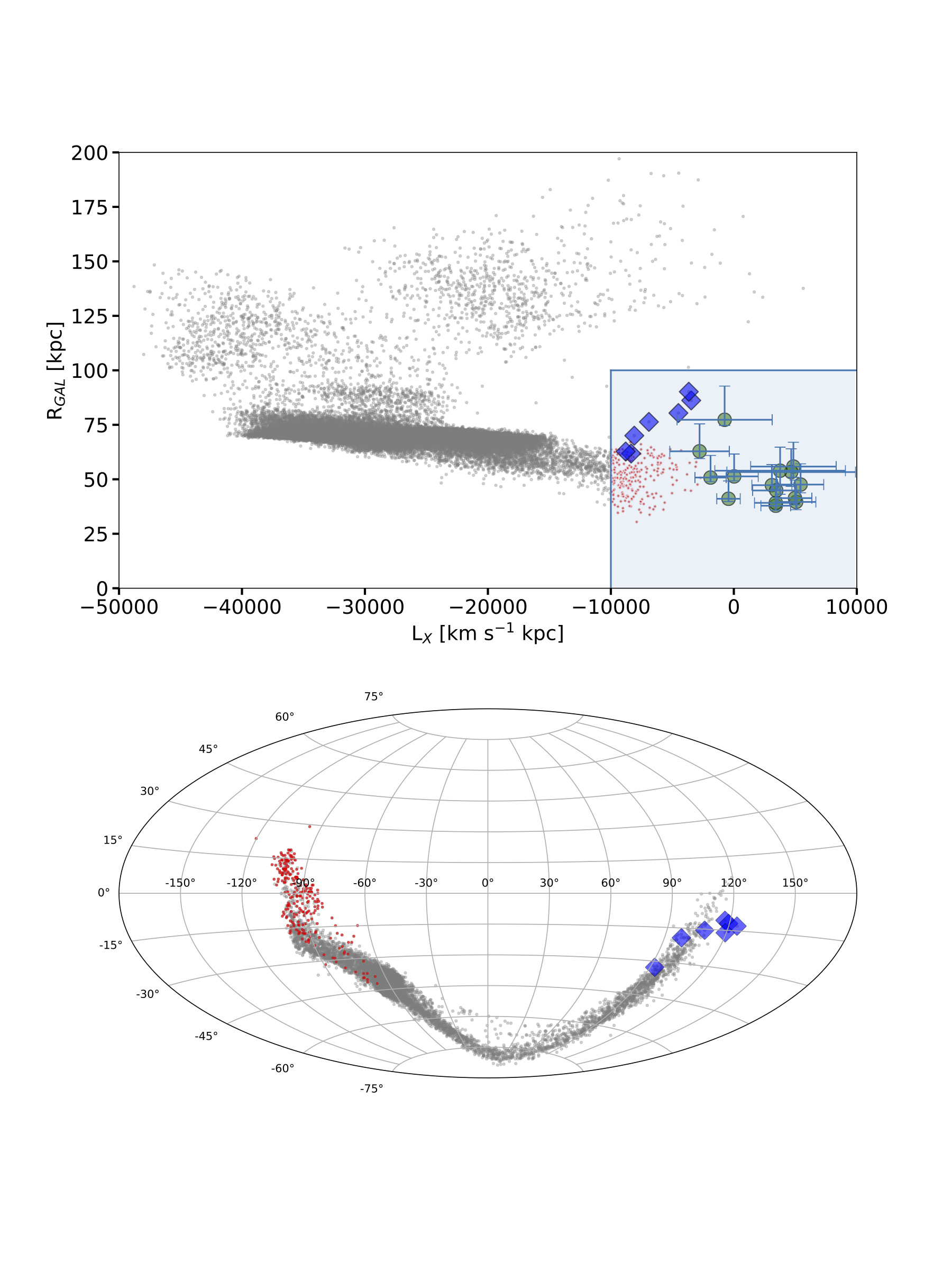}
\caption{Comparison to numerical simulations of SMC stellar tidal debris \citep{besla13}. Because the orbit is mainly confined to the y-z plane, we choose to compare the $x$ component of the angular momentum vector, $L_x$, and $R_{GAL}$ of simulated particles to the H3 stars (large green circles). We highlight the population of interest with the shaded box in the upper panel and distinguish those particles in the simulation that lie in the leading arm (small red circles) to those in the trailing arm (large blue diamonds). In the lower panel, we compare the distribution on the sky of the simulated particles using the same symbols. Uncertainties in $R_{GAL}$ account for the potential $\sim$ 20\% systematic underestimation in distance due to the application of the priors. Uncertainties in $L_x$ do not account for this effect. Using the revised distances drives the location of the H3 stars closer to that of the blue diamonds. 
Although the bulk of the simulated particles at $R_{GAL} < 80$ kpc are either in the body of the SMC or the leading arm, the H3 stars are not entirely without analogs.}
\label{fig:models}
\end{figure}

In the upper panel of Figure \ref{fig:models} we show what we discussed above: that simulations do not predict a significant population of stars with the distance or $L_x$ of our H3 stars. On the other hand, the H3 stars appear to be a plausible extension of the simulated distribution and perhaps a modest change in the initial conditions could produce stars that land in the region of the diagram populated by the H3 stars. Furthermore, when we examine where the simulated debris within this region of parameter space falls on the sky, we find that a subset indeed falls near the tail of the trailing tidal feature (bottom panel of Figure \ref{fig:models}). The correspondence is even closer when we use the prior-free distances estimates. As such, we conclude that although the simulations suggest that the H3 stars are not tracing the bulk of the debris, they are plausibly part of the debris field even with their apparently incongruous physical characteristics. 

Extending this plausibility argument in support of the suggested association of these stars with the SMC tidal debris field, we note
that the mean [Fe/H], the initial value of [Fe/H] for each star is derived in our model fitting, of the H3 stars is $-1.4$ with a standard deviation of $0.2$. This mean value and dispersion are both in excellent agreement with the measured SMC stellar abundance \citep[$-1.35\pm0.10$;][]{depropris} and significantly different from the measured mean abundance of LMC stars \citep[$-0.40$;][]{cole}. This agreement may not, however, be definitive. Based on metallicity arguments alone, the stars could also come from the LMC outskirts, which might be expected to be significantly more metal poor than the stars in the main body. The H3 stars are also not distinct from the overall Galactic halo metallicity distribution. 

\section{Conclusions}
\label{sec:conclusions}

Within the current H3 sample of halo stars, we have identified a subset of 15 distant ($R_{GAL} > 37.5$ kpc), fast-infalling ($-$ 350 km s$^{-1} < V_{GSR} < -155$ km s$^{-1}$) stars that are tightly grouped within the H3 survey footprint and include 94\% of the H3 stars with these characteristics that were not previously associated with the Sagittarius dwarf \citep{johnson}. Furthermore, these stars share a location on the sky and velocity with the gaseous Magellanic Stream and nearby compact gas clouds, leading us to conclude that they too are likely part of the tidal debris field resulting from the interaction of the Magellanic Clouds. 

These stars share the chemical abundance range of stars in the SMC, suggesting that they were extracted from the SMC, although other scenarios remain viable. To explore this possibility further, we examined published simulations of the SMC tidal debris \citep{besla13} and find that although the exact nature of these stars is not reproduced in those simulations, the properties are not sufficiently different to lead us to reject the scenario. We do, however, conclude that if these stars are SMC tidal debris they do not trace the main body of that debris, which is expected to lie well beyond $R_{GAL} = 125$ kpc in this area of the sky and be beyond reach for H3. 

Looking forward, there is the possibility of addressing these various issues with models that include the MW hot corona, possibly rotating and magnetized \citep{tepper-garcia}, 
the Magellanic hot corona \citep{lucchini}, and up-to-date MW and MC mass models to simultaneously match both the gaseous and stellar debris properties.
We acknowledge that our current sample of putative Magellanic tidal debris is small, but
the discovery of these stars, and what we hope is eventually a significant enlarging of the sample and improved proper motions from upcoming Gaia releases, will lead to quantitative  improvements in our understanding of the history of our Galaxy and the dynamical history of the Magellanic Clouds.

\acknowledgments
\label{sec:acknowledgments}

DZ thanks Scott Lucchini, David Nidever, and Elena D'Onghia for timely responses to queries that helped clarify certain issues. J.S.S. would like to acknowledge support from the Banting Postdoctoral Fellowship program and the Dunlap Institute for Astronomy \& Astrophysics. Y.S.T. is grateful to be supported by the NASA Hubble Fellowship grant HST-HF2-51425.001 awarded by the Space Telescope Science. 
We thank the Hectochelle operators Chun Ly, ShiAnne Kattner, Perry Berlind, and Mike Calkins, and the CfA and U. Arizona TACs for their continued support of the H3 Survey. Observations reported here were obtained at the MMT Observatory, a joint facility of the Smithsonian Institution and the University of Arizona. This work has made use of data from the European Space Agency (ESA) mission
{\it Gaia} (\url{https://www.cosmos.esa.int/gaia}), processed by the {\it Gaia}
Data Processing and Analysis Consortium (DPAC,
\url{https://www.cosmos.esa.int/web/gaia/dpac/consortium}). Funding for the DPAC
has been provided by national institutions, in particular the institutions
participating in the {\it Gaia} Multilateral Agreement.

\vspace{5mm}
\facilities{MMT (Hectochelle), \emph{Gaia}}

\software{\texttt{IPython \citep{Perez2007}, matplotlib \citep{Hunter2007}, numpy \citep{van2011numpy}, Astropy \citep{astropy:2018}, SciPi \citep{2020SciPy-NMeth},
Gala \citep{gala}}}

\bibliography{ms}{}
\bibliographystyle{aasjournal}

\end{CJK*}
\end{document}